\documentclass[preprint]{aastex}
\newcommand{\dif}{\mathrm{d}}

\shorttitle{Nonexistence of out-of-plane equilibria}
\shortauthors{Wang et al.}

\begin{document}
%
\title{On existence of out-of-plane equilibrium points in restricted three-body problem with oblateness}

\author{Xuefeng Wang, Nan Wu, Liyong Zhou and Bo Xu}
\affil{School of Astronomy and Space Science, Nanjing
	University, Nanjing 210046, China\\
Key Laboratory of Modern
Astronomy and Astrophysics in Ministry of Education, Nanjing
University}
\email{zhouly@nju.edu.cn} 

\begin{abstract}
We analyze in this paper the existence of the ``out-of-plane'' equilibrium points in the restricted three-body problem with oblateness. From the series expansion of the potential function of an oblate asteroid, we show analytically all equilibrium points locate on the orbital plane of primaries and how artificial equilibrium points may arise due to an inappropriate application of the potential function. Using the closed form of the potential of a triaxial ellipsoid, we analytically demonstrate that the gravitational acceleration in $z$-direction is always pointing toward the equatorial plane, thus it could not be balanced out at any value of $z\neq 0$ and the out-of-plane equilibrium points cannot exist. The out-of-plane equilibrium points appear only when additional acceleration other than the gravitation from primaries is taken into account. We suggest that special attention must be paid to the application of the spherical harmonics expansion of potential to find the equilibrium points, especially when these points may be very close to the celestial body.
\end{abstract}


\section{Introduction}
To find the equilibrium points and their stabilities is a key step to understand the structure of phase space of a dynamical system as well as the orbital behaviours. In the restricted three-body problem concerning the motion of a small (celestial or artificial) object in the gravitational field of two massive primaries, the equilibrium points suggest the existence of temporary (unstable) or permanent (stable) periodic trajectories around the primaries, which not only provide explanations of the orbital configuration of celestial bodies but also benefit the trajectory design for space missions.

In the classical restricted three-body problem, it is well-known that there are five equilibrium points, i.e. three collinear points ($L_1, L_2, L_3$) and two triangular points ($L_4, L_5$), all on the orbital plane of the primaries \citep[see e.g.][]{mur99}.
In the perturbed restricted three-body problem, however, 
these so called Lagrangian points ($L_{1\sim 5}$) still exist but with different locations and stabilities. Detailed studies of their dynamical features can be found e.g. in \cite{abou2016,els2016}.
With the effect of radiation pressure, apart from these five coplanar ones, other equilibrium points than $L_1\sim L_5$ are found and analyzed in literatures, e.g. \cite{sim85,rago88,todo94,roma01,ans2017}. When the oblateness of primaries (one or both) is taken into account, the existence of out-of-plane equilibrium points, locating out of the orbital plane of primaries, is also declared \citep{dou06}, and the dynamics of orbits around these new equilibrium points have been discussed in some other literatures.

However, the existence of such out-of-plane equilibrium points in the latter case is doubtful. For example, \cite{zama15} found in their calculation that such additional equilibrium points in the Mars-Phobos system locate beneath the surface of the Martian moon Phobos, making them physically meaningless. Intuitively, at any point out of the orbital plane, each (oblate) primary's gravitational force on the third body has a component along the $z$ direction, and both of them point to the orbital plane 
that in general coincides with the primary's equatorial plane. Not like in the photogravitatioanl system where the radiation pressure offers a force opposite to the gravitation, there is no other force in this ``oblate-primary'' system that can balance out the combined gravitation in $z$ direction. Thus no force balance can be attained only except for on the orbital plane ($z=0$). 

We will check again in this paper where all the equilibrium points locate in the restricted three-body problem with oblateness, and particularly, we will find out whether we should expect to see equilibrium points out of the orbital plane.

In Section \ref{s:j2}, we will briefly review the potential function of an oblate homogenous body and the calculation of the equilibrium points, as well as show how artificial out-of-plane equilibrium points may arise due to an inappropriate employment of the potential function. In Section \ref{s:ellip}, a more general case where one or both of the primaries are triaxial ellipsoids is investigated, and we demonstrate that equilibrium points must locate on the orbital plane because the force balance can only be attained on the same plane. And we discuss the condition for the existence of out-of-plane equilibrium points and draw conclusion in Section \ref{s:con}.

\section{Motion under the truncated potential} \label{s:j2}
\subsection{Background of the potential function}

\begin{figure}[htbp]
	\centering
	\includegraphics[width=8.0cm,height=6cm,angle=0]{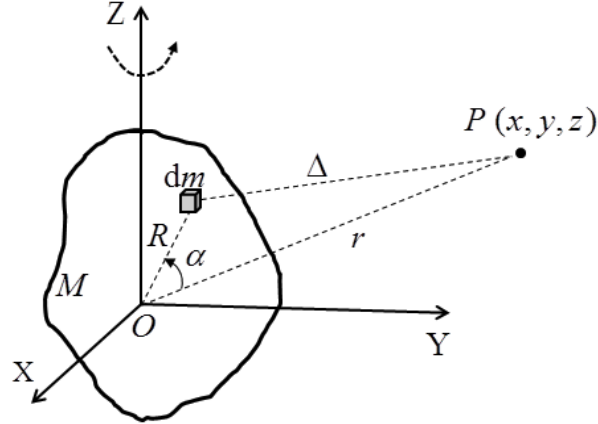}
	\caption{Gravitation potential created by a mass element $\dif m$ on an exterior point $P$. }
	\label{f:1}
\end{figure}

The expansion of a non-spherical body's gravitational potential can be found in many textbooks of celestial mechanics, e.g. \cite{sze67, mur99}. Here we briefly review the perturbation generated by the $J_2$ term for convenience. Assume a homogenous body $M$ of arbitrary shape and a coordinate frame centered in the center of mass, as sketched in Fig.\,\ref{f:1}. The gravitational potential of $M$ at any point $P$ out of the body can be described as an integral 

\begin{equation}\label{e:vori}
V=\int \frac{G \dif m}{\Delta}=\int \frac{G \dif m}{(r^2+R^2-2rR\cos\alpha)^{1/2}}.
\end{equation}
Expand $1/\Delta$ using Legendre polynomials, we get
\begin{eqnarray}\label{e:legexp}
V & = & \frac{G}{r}\int \dif m\sum_{n=0}^{\infty}\left(\frac{R}{r}\right)^n \mathcal{P}_n(\cos\alpha) \nonumber \\
& = & \frac{G}{r}\int \left[ 1+\frac{R}{r}\cos\alpha
 + \frac{1}{2} \left( \frac{R}{r} \right)^2 \left( 3\cos^2\alpha -1 \right) + \cdots\right]\dif m 
\end{eqnarray}
where $\mathcal{P}_n$ is the Legendre polynomials. The $n=1$ term $R\cos\alpha/r$ and $n=2$ term $ (R^2/r^2)\left( 3\cos^2\alpha -1 \right)/2$ respectively are just the $J_1$ and $J_2$ zonal terms in the spherical harmonics expansion of the gravitation potential.
We now simplify the model by supposing that $M$ is a triaxial ellipsoid with three axes $R_x, R_y, R_z$ along the $OX, OY, OZ$ axes and the moments of inertia about these three axes are $I_x, I_y, I_z$. Because the origin of the Cartesian coordinate system $O-XYZ$ coincides with the center of mass, we have
\begin{eqnarray*}
\int\dif m=M, ~~ \int R\cos\alpha \dif m =0, ~~ 
 \int R^2\dif m =\frac{1}{2} \left( I_x + I_y + I_z \right).
\end{eqnarray*}
The potential $V$ now reads
\begin{eqnarray*}
		V= \frac{GM}{r} + \frac{G(I_x+I_y+I_z)}{2r^3} - \frac{3G}{2r^3} \int R^2\sin^2\alpha \dif m + \cdots.
\end{eqnarray*}
From Fig.\,\ref{f:1}, we know that $\int R^2\sin^2\alpha \dif m = I_{OP}$ is the moment of inertia of $M$ about axis $OP$ while the direction cosine of $OP$ is $(x/r,y/r,z/r)$, thus $I_{OP}= \left(I_x x^2 + I_y y^2 + I_z z^2 \right)/r^2$. So now, $V$ becomes
\begin{eqnarray} \label{e:moiexp}
 V  = \frac{GM}{r} + \frac{G(I_x+I_y+I_z)}{2r^3} - \frac{3G\left(I_x x^2 + I_y y^2 + I_z z^2\right)}{2r^5} + \cdots.
\end{eqnarray}

When $M$ is a rotational ellipsoid with equatorial and polar radii $R_e$ and $R_p$,  we have
\[ I_e\equiv I_x=I_y=\frac{1}{5}\left(R_e^2+R_p^2\right)M, ~~ I_p\equiv I_z=\frac{2}{5}R_e^2 M. \]
If $M$ is an oblate ellipsoid, then $R_e > R_p$ and $ I_e < I_p$. Consider the equation:
\begin{eqnarray*}
 I_x x^2 + I_y y^2 + I_z z^2 = I_e(r^2-z^2)+I_pz^2 =  (I_p-I_e)z^2+I_er^2,
\end{eqnarray*}
then $V$ in Eq.\,(\ref{e:moiexp}), truncated up to $J_2$ term, can be rewritten as
\begin{equation}\label{e:trnpot}
V \approx  \frac{GM}{r} + \frac{GM\left(R_e^2 -R_p^2\right)}{10r^3}\left(1- \frac{3z^2}{r^2} \right).
\end{equation}
This potential function is exactly the one widely employed in literatures such as \cite{shar76,obe03,perd05,dou06,sing13b,sing13a,abou15,song16,zen16,zen17}, etc.

One must note that the expansion in Eq.\,(\ref{e:legexp}) only converges when $R<r$, or in other words, the convergence of spherical harmonics series can be guaranteed only outside the Brillouin sphere, which is the smallest sphere enclosing all mass elements \citep[see e.g.][]{romc01}.

\subsection{The equilibrium points in the synodic frame}
Here we will find the equilibrium points in the model of a massless body moving around a non-spherical body with $J_2$ potential term as given in Eq.\,(\ref{e:trnpot}). Then we will extend the analysis to the restricted three-body problem by introducing another point mass in next subsection. 

In this model, an equilibrium point in a rotational frame corresponds to a circular orbit in the inertial coordinate system. In fact, it is well known \citep[see e.g.][]{mur99} that any orbit with nonzero eccentricity and/or nonzero inclination will not be closed thus not correspond to an equilibrium point, because the longitudes of pericentre $\varpi$ and node $\Omega$ will precess due to the $J_2$ perturbation. 
Therefore, we focus only on the circular orbit ($e=0$) on (or parallel to) the equatorial plane ($i=0^\circ$) hereafter. 

\begin{figure}[htbp]
	\centering
    \includegraphics[width=6.0cm,height=6.0cm,angle=0]{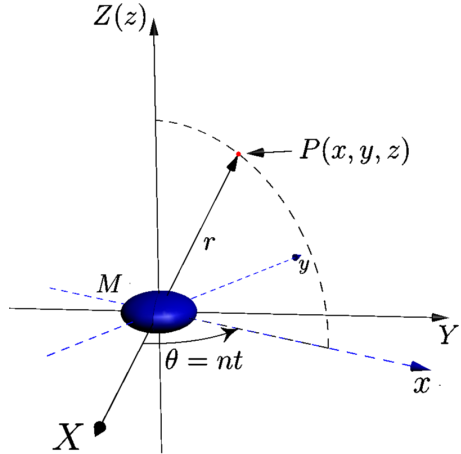}
	\caption{The motion of a massless body under the gravitational attraction from an oblate ellipsoid. $O-XYZ$ is an inertial frame centered at the mass center of $M$ while $O-xyz$ is a rotational frame with angular velocity $n$. }
	\label{f:2}
\end{figure}

In the rotational frame $(O-xyz)$ with angular velocity $n$ (see Fig.\,\ref{f:2}), the equation of motion reads
\begin{eqnarray}\label{e:ros}   
\ddot{x}-2n\dot{y} = \frac{\partial U}{\partial x}, ~~~
\ddot{y}+2n\dot{x} = \frac{\partial U}{\partial y}, ~~~
\ddot{z} = \frac{\partial U}{\partial z}, 
\end{eqnarray}
where $U=\frac{1}{2}n^2(x^2+y^2) + V$ and $V$ is the potential function truncated up to the $J_2$ term, as Eq.\,(\ref{e:trnpot}). The stationary solutions to Eq.\,(\ref{e:ros}) are just the equilibrium points in the rotational frame and circular periodic orbits in the inertial frame. We find all of these solutions in two cases as follows. 

\noindent  \textbf{Case I: $z\equiv 0$}\\
It's easy to obtain the following equations:
\begin{mathletters}
\begin{eqnarray}
n^2x&=&\frac{GM}{r^3}\left[1+\frac{3(R_e^2-R_p^2)}{10r^2}\right]x,  \\
n^2y&=&\frac{GM}{r^3}\left[1+\frac{3(R_e^2-R_p^2)}{10r^2}\right]y. 
\end{eqnarray}
\end{mathletters}
Except for the trivial solution $(x,y,z)=(0,0,0)$, countless solutions can be obtained by solving the equation:
\begin{equation} \label{e:meanmot}
n^2=\frac{GM}{r^3}\left[1+\frac{3(R_e^2-R_p^2)}{10r^2}\right]. 
\end{equation}
For any given $r$ (surely $r>R_e$), there exists an $n=n(r)$ that satisfies the above equation. The equilibrium points (corresponds to periodic circular orbits in inertial frame) are countless on the $O-xy$ plane. 

\noindent \textbf{Case II: $z\neq 0$}\\
The equilibrium equation from Eq.\,(\ref{e:ros}) now reads: 
\begin{mathletters}\label{e:eqzne0}
\begin{eqnarray} 
 n^2x&=&\frac{GM}{r^3}\left[1+\frac{3(R_e^2-R_p^2)}{10r^2}\left(1-\frac{5z^2}{r^2}\right)\right]x,\\
 n^2y&=&\frac{GM}{r^3}\left[1+\frac{3(R_e^2-R_p^2)}{10r^2}\left(1-\frac{5z^2}{r^2}\right)\right]y,\\
 z&=&-\frac{3(R_e^2-R_p^2)}{10r^2}\left(3-\frac{5z^2}{r^2}\right)z.
\end{eqnarray}
\end{mathletters}
Obviously, one solution to Eq.\,(\ref{e:eqzne0}) is
\begin{equation}
x_0=y_0=0, ~~ z_0= \pm\sqrt{\frac{3}{5}\left(R_e^2-R_p^2\right)}.
\end{equation}
At these two points $(x,y,z)=(0,0,z_0)$ on the $Oz$ axis, the gravitation from the main part of the oblate body $M$ is seemingly canceled out by the acceleration generated by the $J_2$ term. However, the distance from this point to the center of $M$ is $r_0=|z_0|<R_e$, i.e. this force balance point in fact is inside the Brillouin sphere, where the expansion adopted above is invalid. Therefore, these equilibrium points are not true. 

For $x\neq 0$ or $y\neq 0$, since $n^2\ge 0$, the terms inside the square brackets in Eq.\,(\ref{e:eqzne0}) must satisfy
\begin{equation} \label{e:12xy}
1+\frac{3(R_e^2-R_p^2)}{10r^2}\left(1-\frac{5z^2}{r^2}\right) \ge 0.
\end{equation}
And from the 3rd equation in Eq.\,(\ref{e:eqzne0}), 
\begin{equation} \label{e:3rdz}
1=-\frac{3(R_e^2-R_p^2)}{10r^2}\left(3-\frac{5z^2}{r^2}\right).
\end{equation}
Substitute Eq.\,(\ref{e:3rdz}) into Eq.\,(\ref{e:12xy}), we get
\begin{equation}
-2\cdot \frac{3(R_e^2-R_p^2)}{10r^2}\ge 0.
\end{equation}
This inequality is impossible for any rotational ellipsoid ($R_e > R_p$). This contradiction indicates that the ``out-of-plane'' equilibrium points with $z \ne 0$ does not exist in this model where the $J_2$ term is taken into account.

\subsection{The equilibrium in restricted three-body problem}

Now we introduce another body (point mass) with mass $m$ on the equatorial plane of $M$ to make a circular restricted three-body problem. Without loss of generality, assume $m$ locates on the $Ox$ axis at a distance $D$ to the first primary $M$. As usual, the distance between primaries $D$ is set as the length unit, the total mass $(M+m)$ as the mass unit, and the time unit is chosen so as to make the gravitational constant $G=1$. Then the oblateness of $M$ can be described by the normalized equatorial and polar radii $r_e=R_e/D$ and $r_p=R_p/D$  respectively. 
After $m$ is introduced, the barycenter of the system is shifted and in the new synodic frame the stationary coordinates of $M$ and $m$ are $(-\mu,0,0)$ and $(1-\mu,0,0)$, where $\mu=m/(m+M)$. And the motion of the massless third body can be described by
\begin{equation} \label{e:ros2}
\ddot{x}-2n\dot{y}=\frac{\partial W}{\partial x},~~~
\ddot{y}+2n\dot{x}=\frac{\partial W}{\partial y},~~~
\ddot{z}=\frac{\partial W}{\partial z},
\end{equation}
where
\begin{eqnarray*}
W&=&\frac{n^2(x^2+y^2)}{2}+\frac{\mu}{r_2}+\frac{1-\mu}{r_1}\left[1+\frac{A}{2r_1^2}\left(1-\frac{3z^2}{r^2_1}\right)\right],\\
r_1&=&\sqrt{(x+\mu)^2+y^2+z^2},~~~ r_2=\sqrt{(x+\mu-1)^2+y^2+z^2},\\
A&=&\frac{R_e^2-R_p^2}{5D^2}=\frac{r_e^2-r_p^2}{5}.
\end{eqnarray*}
Eq.\,(\ref{e:ros2}) is the same as Eq.\,(\ref{e:ros}) only except that the effective potential function $W$ here has the additional potential $\mu/r_2$ from the second primary $m$. Simple calculation shows that the mean motion rate now is $n=\sqrt{1+3A/2}$.

The equilibrium points should be found by solving the equation $\partial W/\partial x=\partial W/\partial y=\partial W/\partial z=0$. Since we are mainly interested in the ``out-of-plane'' equilibrium points in this paper, we focus on the 3rd equation
\begin{equation} \label{e:dirctz}
\frac{\partial W}{\partial z}=-\frac{\mu}{r_2^3}z-\frac{(1-\mu)}{r_1^3}\Big[
1+\frac{3A}{2r_1^2}\left(3-\frac{5z^2}{r_1^2}\right)\Big]z=0.
\end{equation}
$\partial W/\partial z$ is the acceleration along $z$ direction. It's easy to see from Eq.\,(\ref{e:dirctz}) that the acceleration generated by $m$ and the main part of $M$, i.e.
\begin{equation}
-\left(\frac{1-\mu}{r_1^3}+\frac{\mu}{r_2^3}\right)z,
\end{equation}
always points toward the $Oxy$-plane for $z\neq 0$, and only the component generated by the $J_2$ term may point outward from the $Oxy$-plane. When they are balanced, the ``out-of-plane'' equilibrium points appear. However, as we are going to show below, such points are inside the Brillouin sphere, where the expansion of the potential function is invalid.

For $z\neq 0$, since $-\mu/r_2^3$ is definite negative, Eq.\,(\ref{e:dirctz}) can be possibly satisfied if and only if:
\begin{equation}
f(r_1,z)=1+\frac{3A}{2r_1^2}\left(3-\frac{5z^2}{r_1^2}\right)<0.
\end{equation}
Because $z^2\le r_1^2$, the minimum of $f(r_1,z)$ may be attained when $r^2_1=z^2$: 
\[ \min f(r_1,z)= 1+\frac{3A}{2r_1^2}\left(3-\frac{5z^2}{z^2}\right)=1-\frac{3A}{r_1^2}. \]
Thus, for any $(r_1,z)$ of the ``out-of-plane'' equilibrium point, $\min f(r_1,z)\le f(r_1,z)$, i.e.
\begin{equation}
1-\frac{3A}{r_1^2}\le f(r_1,z)<0.
\end{equation}
Consequently, from this inequality an estimation can be obtained for the out-of-plane equilibrium point:
\begin{equation} \label{e:finest}
r_1<\sqrt{3A}=\sqrt{\frac{3(R_e^2-R_p^2)}{5D^2}}<\frac{R_e}{D}.
\end{equation}
Bearing in mind that $D$ is the unit length, Eq.\,(\ref{e:finest}) means that $r_1 < R_e$, i.e. the equilibrium point locates inside the Brillouin sphere around $M$. 

\begin{figure}[htbp]
	\centering
	\includegraphics[width=15cm,height=7.5cm,angle=0]{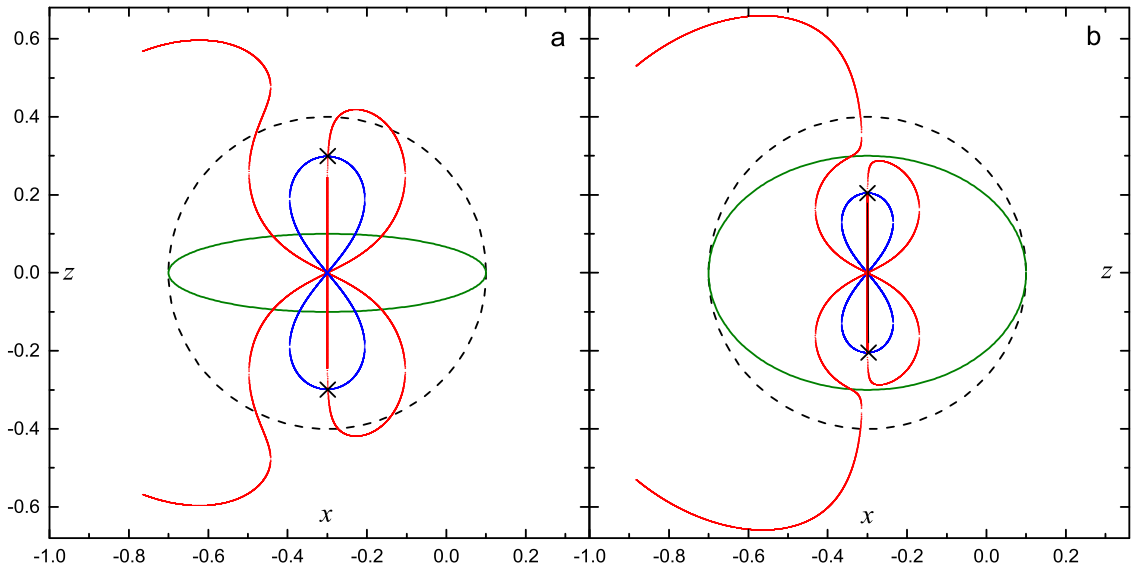}
	\caption{The artificial out-of-plane equilibrium points are inside the Brillouin sphere  or even the ellipsoid. We illustrate the projections on the $Oxz$-plane of the primary ellipsoids (green), Brillouin spheres (dashed black), the contour of surface $\partial W/\partial x=0$ (red) and the contour of surface $\partial W/\partial z=0$ (blue). The equilibrium points at the intersections of red and blue curves are denoted by crosses. }
	\label{f:3}
\end{figure}

In addition, we show in Fig.\,\ref{f:3} numerical calculations of the equilibrium points in two cases with arbitrarily chosen $\mu=0.3$. The oblate primaries in both examples have the same equatorial radius $R_e=0.4$, thus they have the same Brillouin sphere with a radius of $0.4$. For the polar radius, it's $R_p=0.1$ for case (a) and $R_p=0.3$ for case (b). Therefore, the oblateness parameter $A=0.03$ for (a) and $A=0.014$ for (b) respectively. The equilibrium points, denoted by crosses in Fig.\,\ref{f:3}, are clearly inside the Brillouin sphere in case (a) and even inside the ellipsoid surface in case (b) when $A$ is smaller. Apparently, these numerical results support our conclusion drawn above. It is worth to note that we chose relatively large ellipsoid size and parameter $A$ so that the fact can be seen clearly in the figures.

As a matter of fact, these out-of-plane equilibrium points arise only due to the existence of acceleration pointing outward from the orbital (equatorial) plane, and we know now that for an oblate primary body such acceleration component is an artificial effect of the improper application of the potential series of spherical harmonics. We can also show that this conclusion holds if higher order terms $J_{2n}$ are included. We are going to show in next Section that there is no such outward acceleration at all using the closed form of gravitational potential of a triaxial ellipsoid model.

However, if the asymmetry of the gravity (e.g. $J_3$ or tesseral harmonic terms) is considered, there will be the acceleration component pointing outward from the $Oxy$-plane, and the classical equilibrium points $L_{1-5}$ perhaps will not stay on the same plane.
But in this case, we cannot assume that the two primaries will move on a circular orbit on the primaries' equatorial plane, and this will make a totally different model, out of the scope of this paper.

\section{Potential of a triaxial ellipsoid} \label{s:ellip}
A triaxial ellipsoid is a good model for describing an asteroid if it is not too much distorted, and it severs as one of the basic models in many studies. The gravitational potential of a triaxial ellipsoid will be given by an integral in this Section, so that no artificial effects may be introduced due to the truncation of series expansion. We will show that the acceleration arising from a triaxial ellipsoid is always pointing to the equatorial plane, so that there will be no chance for the force balance out of the equatorial plane. Therefore, all equilibrium points must be on the equatorial plane and the out-of-plane equilibrium points do not exist.

In a frame of reference rotating with the ellipsoid, the ellipsoid has a surface defined by
\begin{equation} \label{e:ellipsoid}
\frac{X^2}{A^2}+\frac{Y^2}{B^2}+\frac{Z^2}{C^2}=1,
\end{equation}
where $A\leq B \leq C$ are the long, intermediate, and small semi-axes of the ellipsoid. Apparently, these axes rest exactly on the coordinate axes. After normalizing the coordinates by $A$,
the semi-axes read
\[ a\equiv 1,~~ b\equiv \frac{B}{A}, ~~ c\equiv \frac{C}{A}. \]
And the gravitational potential outside the ellipsoid is \citep[see e.g.][]{sch94,rom12}
\begin{eqnarray} \label{e:potent}
V(x,y,z) =
\frac{3GM}{4A^3}\int_u^{\infty}\left( 1-\frac{x^2}{a^2+s}-\frac{y^2}{b^2+s} 
  -\frac{z^2}{c^2+s} \right)
\times \frac{\dif s}{\sqrt{(a^2+s)(b^2+s)(c^2+s)}},
\end{eqnarray}
where $u$ is a positive root of the following equation
\begin{equation}\label{e:eofr}
\frac{x^2}{a^2+u}+\frac{y^2}{b^2+u}+\frac{z^2}{c^2+u}=1.
\end{equation}
Therefore, the acceleration in the $Oz$ direction can be calculated \citep{rom12}
\begin{eqnarray} \label{e:acceli}
\frac{\partial V}{\partial z}= -\frac{3GMz}{2A^3}\int_u^{\infty}\frac{\dif s}{(c^2+s)\sqrt{(a^2+s)(b^2+s)(c^2+s)}}.
\end{eqnarray}
Because the integral in Eq.\,(\ref{e:acceli}) is definite positive, we have
\begin{eqnarray*} \label{e:fdir}
&\frac{\partial V}{\partial z}   < 0 & \mbox{ for all } z>0, \\
&\frac{\partial V}{\partial z}   = 0 & \mbox{ for    } z=0, \\
&\frac{\partial V}{\partial z}   > 0 & \mbox{ for all } z<0.
\end{eqnarray*}
The acceleration always points toward the $Oxy$-plane, and the acceleration balance in $z$-direction can be attained only when $z=0$. The out-of-plane equilibrium points do not exist.

\section{Conclusion} \label{s:con}

We have shown analytically that in the restricted three-body problem with one oblate primary the ``out-of-plane'' equilibrium points do not exist. The same arguments can be carried out and the same conclusion can be made if both primaries are oblate. The existence of such ``out-of-plane'' equilibrium points is due to an inappropriate application of the potential series of spherical harmonics, which gives rise to a fake vertical acceleration pointing outward the orbital plane.

Surely, when the outward vertical acceleration does exist in a system, the out-of-plane equilibrium points may appear. For example, when one of the primaries are very luminous, the radiation pressure  may overcome the gravitation from it, then such kind of new equilibrium points may appear. This is possible for dust particles or spacecrafts with very large area-mass-ratio $(A/m)$, like solar sails. Another example is the artificial equilibrium points created by low thrust. No matter in which case, we suggest that special attention should be paid to the application of the spherical harmonic expansion in the close vicinity of a celestial body.

Last but not least, the argument presented in this paper hardly affects the calculation of location and stability of the classical equilibrium points, $L_1\sim L_5$, provided that they are not too close to the oblate body.

\acknowledgments
We thank the National Natural Science Foundation of China (NSFC, Grants No.11473016 \& No.11333002) for financial support.

\end{document}